# Design and Analysis of Pseudospin-Polarized Ultra-Wideband Waveguide Supporting Hybrid Spoof Surface Plasmon Polaritons

KAZEM ZAFARI,[1] ATEFEH ASHRAFIAN,[2],[*] MOHAMMAD PASDARI-KIA,[3] HAMED SAGHAEI,[4] MAHDI NOOSHYAR,[5] AMIRMASOOD BAGHERI,[3]

1. Iran University of Science and Technology
2. University of Tehran
3. Sharif University of Technology
4. Islamic Azad University Shahrekord brach
4. University of Mohaghegh Ardabili

**Abstract:** In this study, novel low-loss waveguides and power dividers for ultra-broadband surface plasmon polaritons (SPPs) are introduced. This article uses complementary metasurfaces in place of traditional SPP, which are typically produced as metasurface unit cells on the dielectric sublayer. It has been demonstrated that the use of complementary metasurfaces considerably improves wave confinement and inhibits wave propagation. Because of this, it is anticipated that waveguides and power dividers made from these complimentary unit cells will have significantly lower losses and a greater bandwidth than SPP used in traditional devices. In the frequency range of 0–100 GHz, waveguides and bent waveguides with complementary metasurface unit cells exhibit S21>–0.5dB. Utilizing complementary metasurfaces, symmetric and asymmetric power dividers have been created and researched. The results of the simulation have shown that using this type of unit cell in the construction of microwave devices is advantageous.

## 1. Introduction

The fascinating field of electromagnetic engineering has recently received a lot of attention because of the outstanding abilities of complementary metasurfaces to confine electromagnetic energy within smaller scales spanning the visible and near-infrared spectrums [1, 2]. These metasurfaces have made it possible to downsize photonic circuits and interconnects by promoting the generation of surface plasmon polaritons (SPPs) [3-6]. The advantages of SPPs in the optical spectrum are constrained at lower frequencies, particularly in the microwave and terahertz ranges. The hunt for novel solutions sparked by this discovery gave rise to the concept of spoof surface plasmon polaritons (SSPPs) [7, 8].

A wide range of technical improvements have been made possible by the rapid expansion of the application potential of SSPPs in the microwave and terahertz domains [9,10]. Now that the corresponding plasma frequency can be tailored, SSPPs can be introduced in previously unexplored frequency ranges thanks to cleverly engineered periodic structures on metallic surfaces [11, 12]. In addition to allowing for more effective electromagnetic wave penetration through metals, these finely structured surfaces adorned with subwavelength structures also confine these waves at smaller scales [9]. This innovation opened the door for the creation of a diverse range of SSPP-based devices, each of which offers distinct electromagnetic functions [13-14]. The integration of SSPPs with the field of planar microwave circuits, which has attracted significant interest [15, 16], is of noteworthy significance. SSPP transmission lines have become a major area of research because of their extensive frequency coverage and capacity to precisely manage cutoff frequencies through customized designs [17, 18].

The multitude of SSPP-based gadgets that have appeared is evidence of the development of SSPPs from theoretical notions to practical developments. These components, which range from antennas to power dividers and filters, highlight the tremendous ability of SSPPs to fundamentally alter microwave circuits [19,20]. Novel SSPP waveguides and symmetric/asymmetric dividers stand out in particular for their revolutionary functions, decreased loss, and ultra-wideband capabilities, which enable performance never before seen [21-19].

Additionally, a new aspect of electromagnetic mode control has been added by the incorporation of complementary metasurfaces into the SSPP framework [18-20]. These metasurfaces, which enable increased electromagnetic mode manipulation and seamless integration with traditional microwave devices by modifying unit cell dimensions and configurations, integrate with SSPPs in a seamless manner [19-21]. Recently, a novel electromagnetic phenomenon has emerged known as the "line wave" (LW), rooted in the concept of complementary impedance boundaries [22]. This innovative wave mode is generated by enforcing inversion symmetries at the interface of two complementary metasurfaces, resulting in a pseudospin-polarized line mode. The LW showcases remarkable unidirectional propagation capabilities and operates across a wide bandwidth. Many waveguides and dividers have also leveraged the use of complementary metasurfaces, capitalizing on the unique properties of the LW [23-26].

In this study, SSPP waveguides and splitters were constructed using the potential of complementary metasurfaces. The implementation of this specific unit cell design leads to decreased losses, increased confinement, and higher beamwidth for SSPP structures, demonstrating the effectiveness of the novel approach used. The article is organized as follows: The investigation and design of the waveguide employing the complementary unit cell are covered in the first section. The bent waveguide is described and examined in the second section. Different waveguide topologies are introduced in the third section in accordance with how the metasurface unit cells are organized, and in the fourth section, the design and analysis of symmetric and asymmetric power dividers are taken into consideration. Finally, conclusions are formed.

.
## 2. Waveguide design

We employ the pseudospin states theorem in an effort to design a new ultra-wideband wavegu ide. According to the theorem of pseudospin states, complementary structures are required to design one-way waveguides which sustain spin-like decoupled modes [26]. Here, we establish mirror reflection symmetries by complementary structures that support SPPs. Complementary metasurface unit cell structure is mounted on a Rogers RT5880 dielectric substrate with $\varepsilon_r=$

2.2 dielectric constant. Its geometrical characteristics are $P_x=1$, $P_y=0.22$mm, $a=0.19$mm, $b=0.17$ mm, $c=0.03$mm, and $h=0.15$mm. Figure 1(b) shows the dispersion curves for the complimentary modes, the upper surface only mode, and the lower surface only mode. These curves were computed using the CST software's Eigenmode solver. The complementary unit cell has a larger phase speed than the situations with only one of the levels, either the top level or the lower level that reflects the common SPP, as seen by the dispersion curves. This indicates that the unit cell that was introduced is more confined than the other two examples. Therefore, it is a good idea to use this unit cell when creating waveguides and diodes. It also goes without saying that using this unit cell significantly lowers transmission loss, a topic that will be covered further.

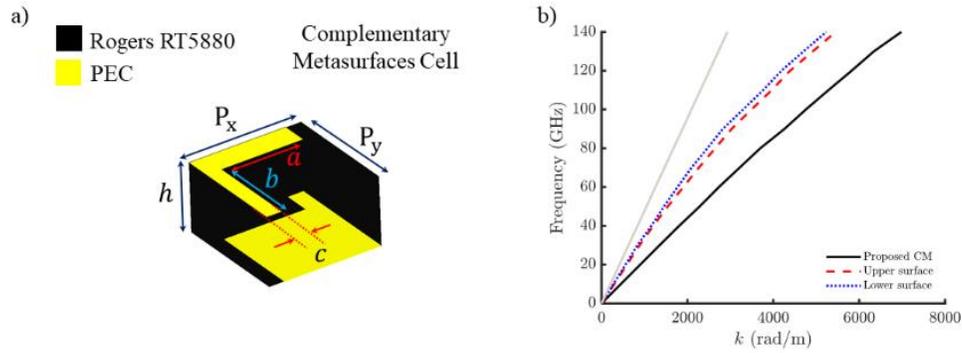

Fig. 1. (a) Schematic of a unit cell on a Rogers 5880 dielectric substrate with the geometric parameters $P_x=1$, $P_y=1$mm, $a=0.1$mm, $b=0.1$, $c=0.05$mm, and $h=0.15$mm. (b) Dispersion diagram for the complementary unit cell as well as for the scenario in which just one of the top or lower surfaces is on the substrate.

A waveguide with dimensions of L = 6.55 mm and W = 5 mm is created using the newly introduced unit cell, as seen in Figure 2(a). Additionally, Figure 2(b) depicts the electric fields in the XY and ZX planes for the introduced waveguide at 25 GHz and 50 GHz. As shown in Figure 2(a) complementary structures satisfy mirror reflection symmetries about the x-y plane. The hall effect and quantum hall effect occur when a magnetic field breaks the time-reversal symmetry of a system. In the hall effect, an electric field is created that is perpendicular to both current flow and the magnetic field applied. On the other hand, the spin hall effect of light (SHEL) is the photonic equivalent of the quantum spin hall effect, where the direction of momentum defines the unique intrinsic spin of electrons. To demonstrate the SHEL, we utilize a spin source consisting of two orthogonal dipoles with a specific phase difference of $\pi/2$. Placing a point source at a distance less than the wavelength above the structure generates longitudinal spin angular momentum. The waveguide restricts the incident wave's opposite helicities, resulting in unidirectional waves propagating in opposite directions. Figures 2(c), and 2(d) demonstrate the unidirectional propagation of pseudospin states in which the spin-momentum locking of the decoupled modes is evident. This configuration of one-way modes, combined with magneto-optical scattering or absorption, enables efficient optical diodes.

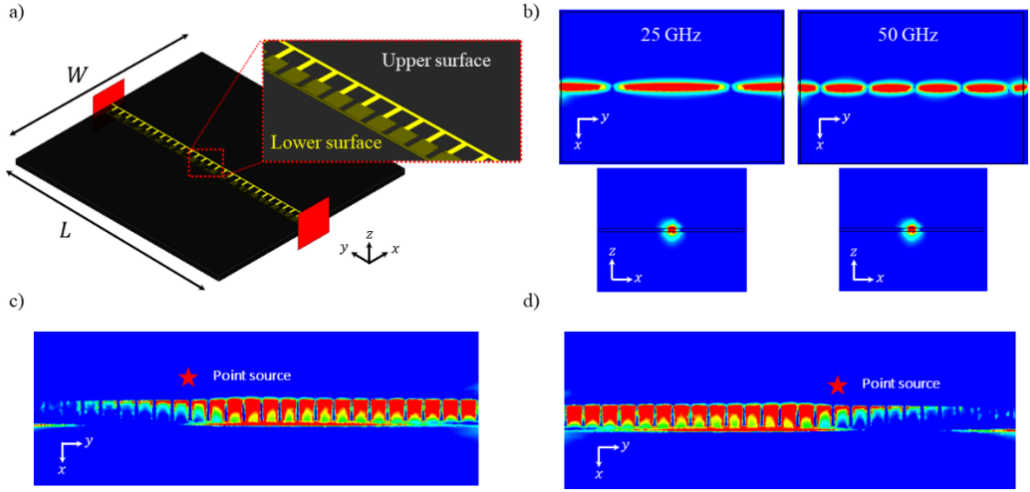

Fig. 2. (a) Schematic of the introduced waveguide using the complementary unit cell, (b) electric field distributions in 25 GHz and 50 GHz in XY plane, (c) , and (d) directional excitation of pseudospin-up and pseudospin-down states using a spin source.

$S_{11}$ and $S_{21}$ curves in the frequency range of 0-150 GHz can be seen in Figures 3(b) and 3(a). Figure 3(b) shows that the value of $S_{21}$ is >-0.5 dB in the frequency range of 0–100 GHz. It is also evident that the losses considerably increase if only one level (the common SPP) is involved. It should be emphasized that if one of the upper metal surfaces—or its complement—is present on the bottom layer, this is truly a typical structure that has been examined in a variety of sources.

Assuming that the structure's electric field has the following composition: $E = E_0 e^{-\alpha_z z} e^{-jk_y y}$, where $\alpha_z$ is the attenuation coefficient in the direction of the vertical axis z and $k_y$ is the propagation constant in the direction of the y axis. For the three complementary unit cell scenarios of the newly introduced waveguide size $\frac{|E|}{|E_0|}$ in Figure 3(c), just the top layer and only the bottom layer are depicted along the x and y = 1 mm axes. It is evident that the complementary unit cell has a larger field attenuation. because as you move away from the field, it falls more forcefully. For this waveguide, using the form of the fields extracted from CST software, the values for the three states have been calculated and plotted in Figure 3(d). It can be seen that the value of the attenuation coefficient for the complementary unit cell is higher than the other two modes, which means more confinement of the wave in the complementary structure. Therefore, it is expected that, due to the increased confinement, the waveguide performance of the introduced structure will be better than the other modes.

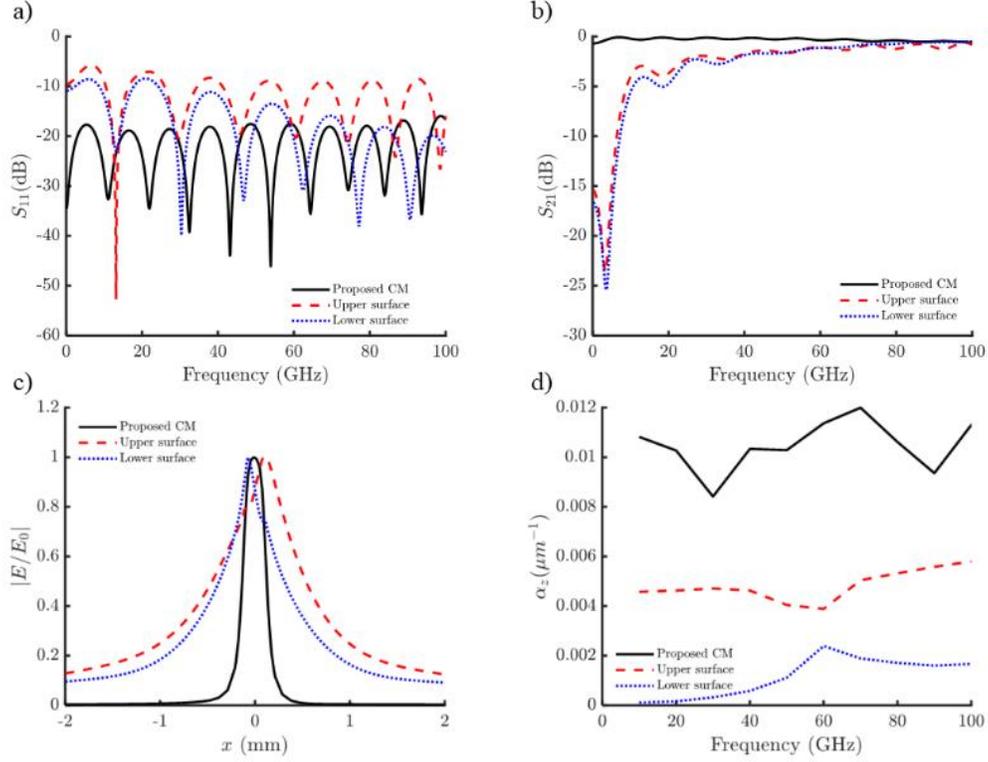

Fig. 3. (a) $S_{11}$ and (b) $S_{21}$ parameters related to the waveguide when the sole unit is complementary and when only one of the surfaces is on the substrate. (c) Magnitude of the normalized electric field along the X-axis for three complementary unit cell states, top surface and bottom surface. (d) Attenuation coefficient $\alpha_z$ for three introduced modes.

## 3. Bent Waveguide

In this section, the complementary unit cell, whose structure is depicted in Figure 4(a), is used to analyze the bent waveguide. The electric field in the XY plane for the frequency of 50 GHz is shown for this structure, as are the states when just one of the metasurfaces is on the substrate. This field distribution demonstrates that the complimentary unit cell confinement is stronger than that in the other two models, preventing the wave from propagating over space. The parameters $S_{11}$ and $S_{21}$ are displayed for this structure in Figures 4(c) and 4(d). As can be shown, the introduced bent waveguide has much less loss than the other two modes for the frequency range of 0-100 GHz, which is less than -1 dB.

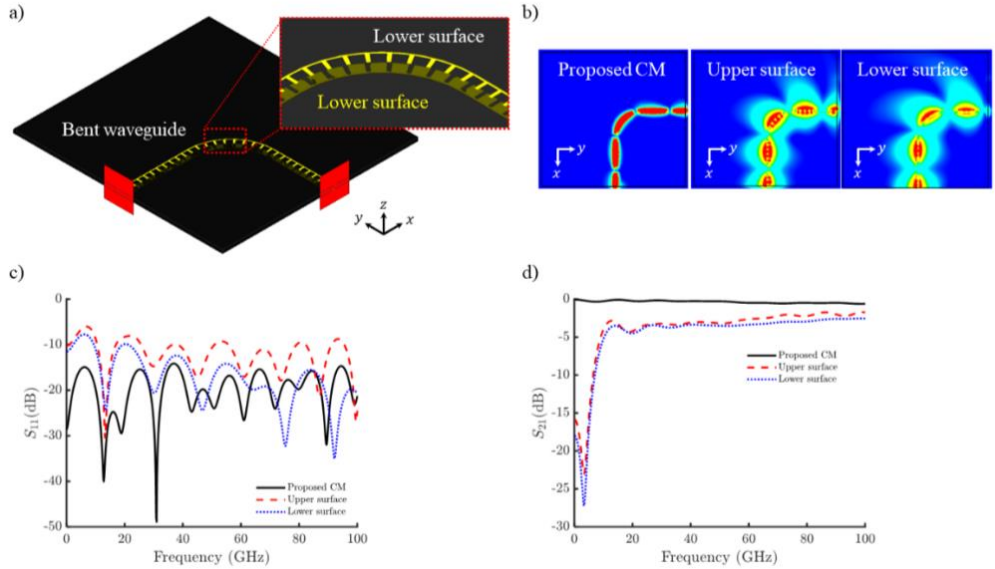

Fig. 4. (a) bent waveguide by using the complementary unit cell. (b) magnitude of the electric field at 50 GHz, (c) $S_{11}$ parameter and (d) $S_{21}$ parameter for the bent waveguide in cases where the unit cell is used complementary, or only in the lower layer or only in the upper layer

## 4. Other different waveguide topologies

In this section, we look at four distinct waveguide topologies utilizing the unit cell that is provided. Figure 5(a) depicts Waveguide I, which consists of the bottom dielectric layer's complementary surfaces that are arranged side by side. Waveguide II is produced in Figure 5(b) by matching upper and lower surfaces. Waveguides III and IV, which lack a complementary unit cell in their construction, are depicted in Figures 5(c) and 5(d).

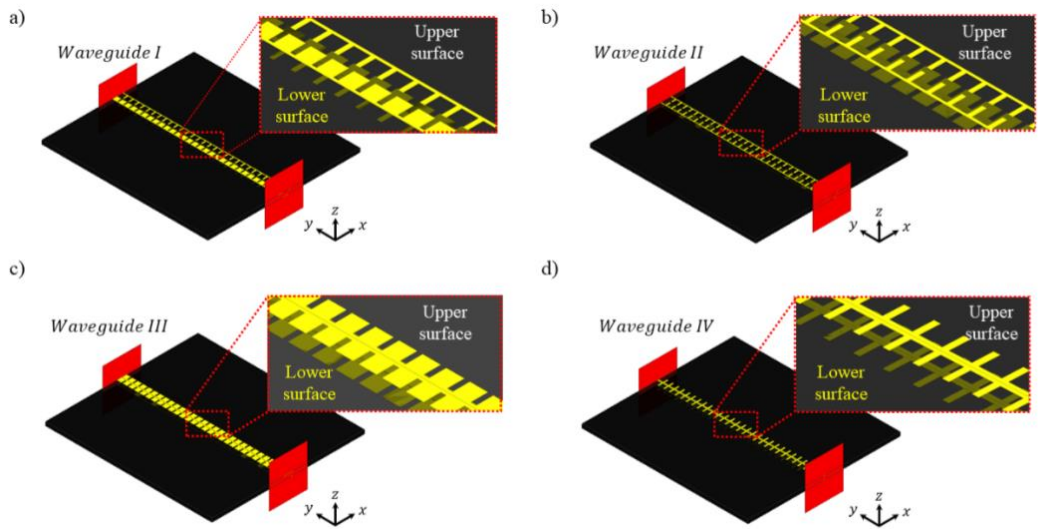

Fig. 5. Schematic of waveguide (a) I, (b) II, (c) III and (d) IV.

The S11 reflection coefficient parameter for waveguides I, II, III, and IV is shown in Figure 6(a). The $S_{21}$ parameter and the dispersion curves that were derived using the Eigenmode solver in the CST software are also reported in Figures 6(b) and 6(c), respectively. The attenuation coefficient is displayed in Figure 6(d) for the four newly introduced waveguides I, II, III, and IV. The observed results lead to the conclusion that waveguides I and II perform better than structures III and IV.

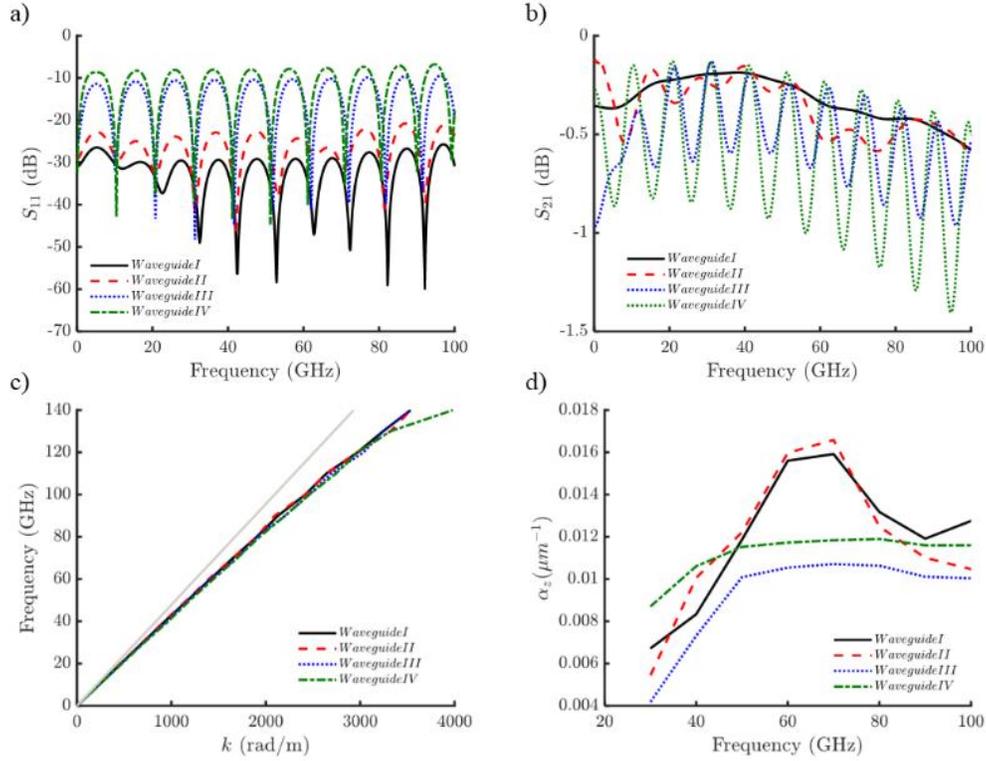

Fig. 6. (a) S11 parameters, (b) S21 parameter, (c) dispersion diagram and (d) attenuation constant $\alpha_z$ related to waveguide I, II, III and IV.

## 5. Power divider design

This section has examined the design of the power divider utilizing the four waveguides I, II, III, and IV shown in Figure 5. Four power dividers—Divider I, Divider II, Divider III, and Divider IV—presented in Figure 7 are equivalent to the previously introduced waveguides.

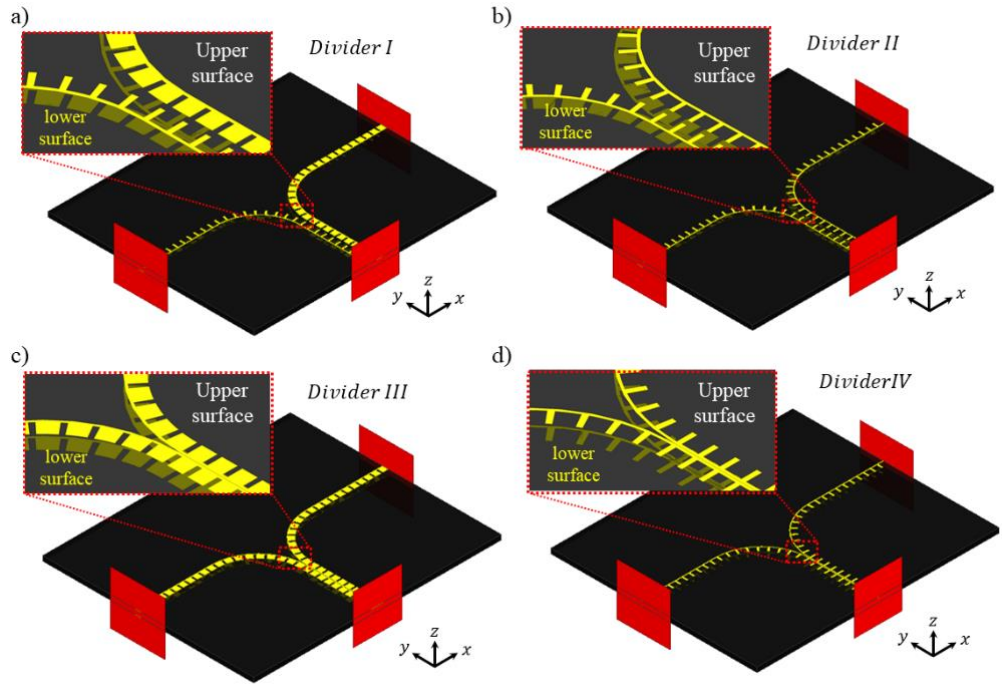

Fig. 7. Schematic of power divider (a) I, (b) II, (c) III and (d) IV.

Parameters $S_{11}$, $S_{21}$, $S_{31}$ and $S_{23}$ for four power dividers—Divider I, Divider II, Divider III and Divider IV—are drawn in Figures 8(a)-(d). The noteworthy point is that Divider I is an asymmetric power divider and the power and phase are different for ports 2 and 3. For this power divider, the value of $S_{21} < -4$ dB and the value of $-3$ dB $< S_{31} < -2.5$ dB. It can also be seen that the isolation between the output ports for power dividers I and II is less than -10 dB ($S_{23} < -10$ dB) while for power dividers III and IV it is -6 dB $< S_{23} < -4$ dB.

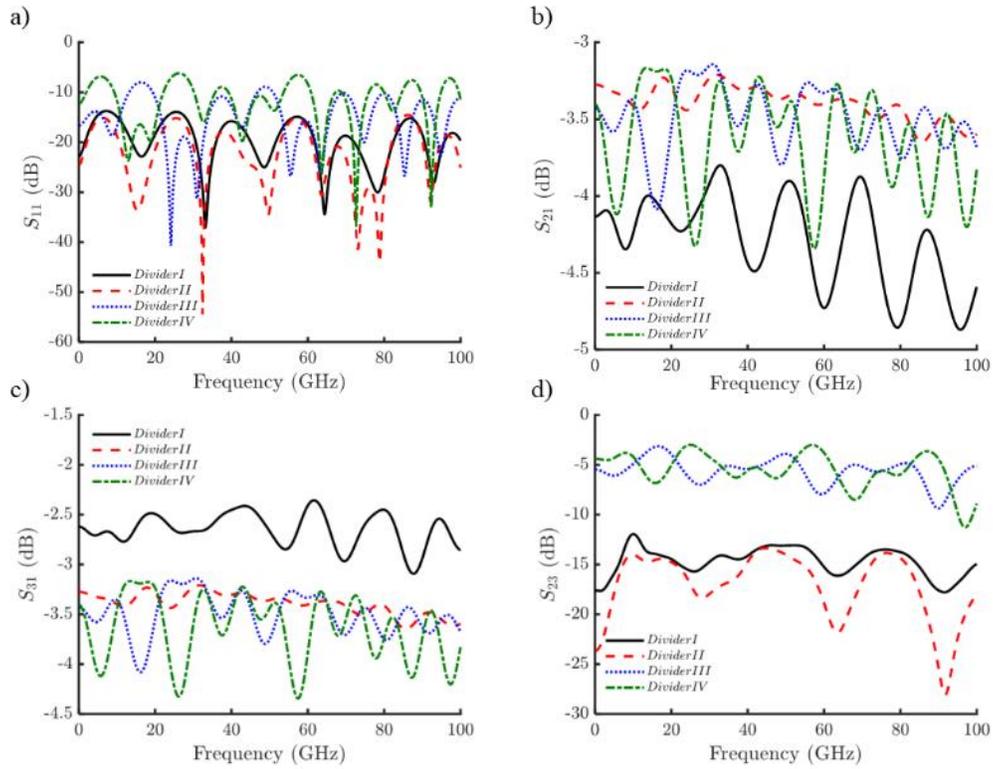

Fig. 8. (a) $S_{11}$, (b) $S_{21}$, (c) $S_{31}$ and (d) $S_{23}$ parameters related to power dividers I, II, III, IV.

Figure 9 shows the phase difference value between ports 2 and 3 for Divider I, Divider II, Divider III, and Divider IV. As can be observed, this difference is approximately zero for the power divider, whereas it is 180 degrees for the power divider I's output ports.

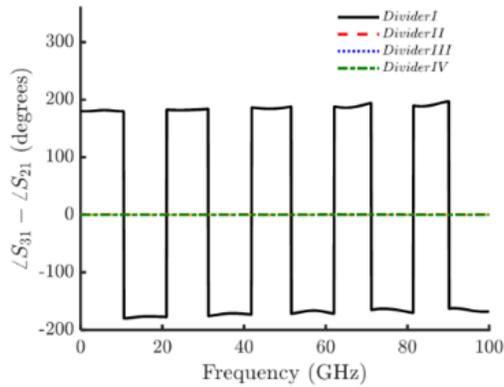

Fig. 9. Phase difference in output ports 2 and 3 for introduced power dividers I, II, III, IV.

Three structures are shown in Figure 10(a)–(c) to compare the performance of the power divider with modes where only one of the upper or lower levels is available.

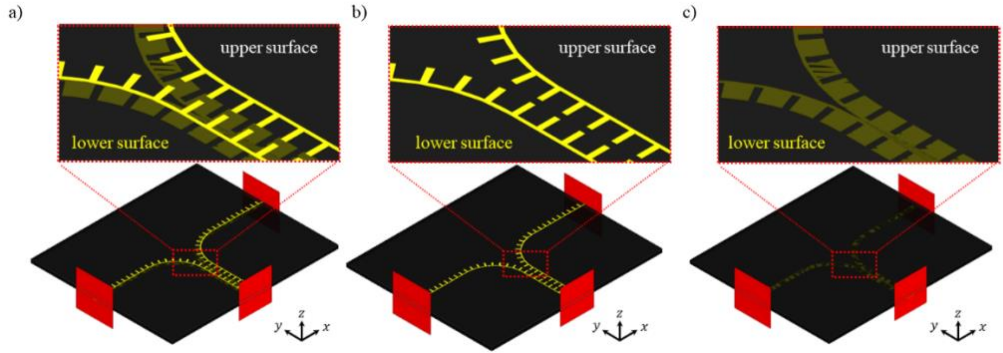

Fig. 10. The power divider consists of: (a) complementary unit cell, (b) top layer and (c) bottom layer.

In Figures 11(a) and 11(b), the $S_{11}$ and $S_{21}$ parameters related to the three structures presented in Figure 10 are reported. It can be seen that using a topology consisting of complementary metasurfaces greatly improves the performance of the structure. So for power dividers, if it is only one of the levels (the common SPP), the value of $S_{21}$ is <-5 dB.

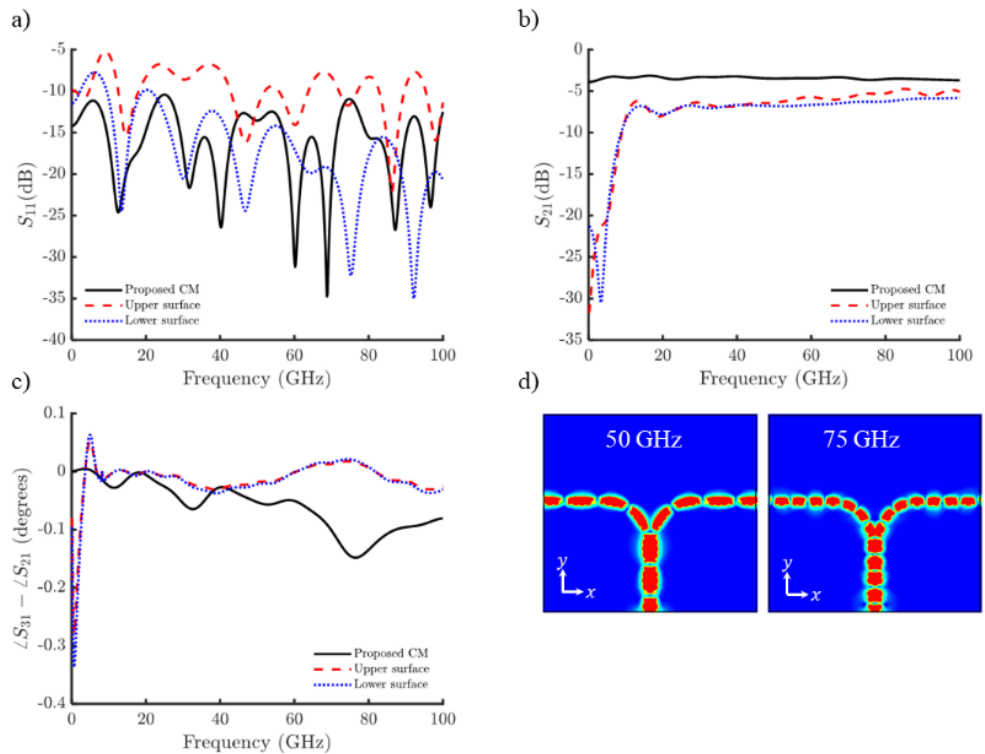

Fig. 11. (a) $S_{11}$ parameters, (b) $S_{21}$ parameters and (c) phase difference between port 2 and 3 for the power divider in cases where the unit cell is used complementary, or only in the lower layer or only in the upper layer. (d) magnitude of electric field in 50 GHz and 75 GHz frequencies.

A power divider with one of its arms inverted is depicted in Figure 12. It appears as though the layer's upper and lower layers have been moved. $S_{11}$, $S_{21}$, and $S_{31}$ characteristics are shown in

Figure 12(b). It is evident that this modification results in filtering in the associated arm. The value of $S_{31} < -10$ dB confirms the issue of filtering. The size of the electric field distribution in the ZY plane for the 25 GHz and 50 GHz frequencies is given in Figure 12(c-d).

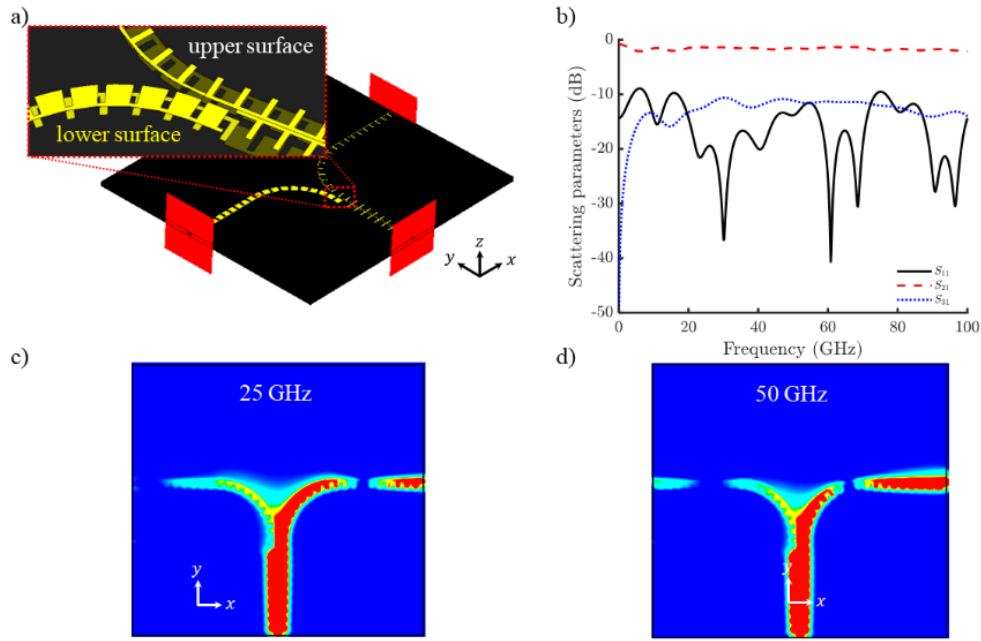

Fig. 12. (a) Schematic and (b) $S_{11}$, $S_{21}$ and $S_{31}$ parameters of the power divider for the case where one of the branches is reversed. (c) magnitude of electric field in 25 GHz and (d) 50 GHz frequencies.

## 6. Conclusion

This work introduces a series of complementary metasurface-based very wideband waveguides and power dividers for the frequency range of 0-100 GHz. The introduction of complementary unit cells considerably enhances performance, as this paper's comparison of their use with traditional SPPs demonstrates. The attenuation coefficients for the modulated unit cell and supplemental unit cell (if one of the levels exists) have also been provided in this article, and it has been demonstrated that using the supplementary unit cell increases both confinement and attenuation coefficient.

**References**

1. W. L. Barnes, A. Dereux, and T. W. Ebbesen, "Surface plasmon subwavelength optics," Nature, vol. 424, pp. 824–830, Aug. 2003

2. Maier, Stefan A. *Plasmonics: fundamentals and applications*. Vol. 1. New York: springer, 2007.
3. Shen, Xiaopeng, et al. "Conformal surface plasmons propagating on ultrathin and flexible films." *Proceedings of the National Academy of Sciences* 110.1 (2013): 40-45.
4. Ahmadi, Haddi, Zahra Ahmadi, Nasrin Razmjooei, Mohammad Pasdari-Kia, Amirmasood Bagheri, Hamed Saghaei, Kamalodin Arik, and Homayoon Oraizi. "Line-wave waveguide engineering using Hermitian and non-Hermitian metasurfaces." *Scientific Reports* 14, no. 1 (2024): 5704.
5. Di Paola, Antonio, Massimo Moccia, Giuseppe Castaldi, and Vincenzo Galdi. "Coupled line waves in parallel-plate metasurface waveguides." *MRS Communications* (2024): 1-7.
6. Ahmadi, H., Ahmadi, Z., Karimzadeh, R., Saghaei, H., Ghoreishi, F. S., Zafari, K., ... & Bagheri, A. (2024). Line waves Supported by Non-Hermitian Metasurfaces Satisfying Electromagnetic Duality.
7. Törmä, Päivi, and William L. Barnes. "Strong coupling between surface plasmon polaritons and emitters: a review." *Reports on Progress in Physics* 78.1 (2014): 013901.
8. Zhang, Hao Chi, et al. "Smaller-loss planar SPP transmission line than conventional microstrip in microwave frequencies." *Scientific reports* 6.1 (2016): 23396.
9. Kianinejad, Amin, et al. "Spoof plasmon-based slow-wave excitation of dielectric resonator antennas." *IEEE Transactions on Antennas and Propagation* 64.6 (2016): 2094-2099.
10. Wang, Jun, et al. "Splitting spoof surface plasmon polaritons to different directions with high efficiency in ultra-wideband frequencies." *Optics Letters* 44.13 (2019): 3374-3377.
11. Zhou, Shi-Yan, et al. "Four-way spoof surface plasmon polaritons splitter/combiner." *IEEE microwave and wireless components letters* 29.2 (2019): 98-100.
12. Chen, Feng-Jun, et al. "A four-way microstrip filtering power divider with frequency-dependent couplings." *IEEE Transactions on Microwave Theory and Techniques* 63.10 (2015): 3494-3504.
13. Zhou, Yong Jin, and Qian Xun Xiao. "Electronically controlled rejections of spoof surface plasmons polaritons." *Journal of Applied Physics* 121.12 (2017).
14. Zhang, Hao Chi, et al. "Broadband amplification of spoof surface plasmon polaritons at microwave frequencies." *Laser & Photonics Reviews* 9.1 (2015): 83-90.
15. Liu, Shuo, et al. "Anisotropic coding metamaterials and their powerful manipulation of differently polarized terahertz waves." *Light: Science & Applications* 5.5 (2016): e16076-e16076.
16. Unutmaz, Muhammed Abdullah, and Mehmet Unlu. "Terahertz spoof surface plasmon polariton waveguides: a comprehensive model with experimental verification." *Scientific reports* 9.1 (2019): 7616.
17. Chen, Yongyao, et al. "Effective surface plasmon polaritons on the metal wire with arrays of subwavelength grooves." *Optics Express* 14.26 (2006): 13021-13029.
18. Gao, Li-Hua, et al. "Broadband diffusion of terahertz waves by multi-bit coding metasurfaces." *Light: Science & Applications* 4.9 (2015): e324-e324.
19. Ren, Bocong, et al. "Compact spoof surface plasmonic waveguide with controllable cutoff frequency and wide stop band." *Applied Physics Express* 14.2 (2021): 024002.
20. Pan, Bai Cao, et al. "Wideband miniaturized design of complementary spoof surface plasmon polaritons waveguide based on interdigital structures." *Scientific Reports* 10.1 (2020): 3258.
21. Maier, Stefan A., and Stefan A. Maier. "Surface plasmon polaritons at metal/insulator interfaces." *Plasmonics: Fundamentals and Applications* (2007): 21-37.
22. Zhang, Yawei, et al. "A Mach–Zehnder Interferometer Refractive Index Sensor on a Spoof Surface Plasmon Polariton Waveguide." *Electronics* 11.23 (2022): 3944.
23. Jones, Andrew C., et al. "Mid-IR plasmonics: near-field imaging of coherent plasmon modes of silver nanowires." *Nano letters* 9.7 (2009): 2553-2558.

24. Kamaraju, Natarajan, et al. "Subcycle control of terahertz waveform polarization using all-optically induced transient metamaterials." *Light: Science & Applications* 3.2 (2014): e155-e155.
25. Dia'aaldin, J. Bisharat, and Daniel F. Sievenpiper. "Guiding waves along an infinitesimal line between impedance surfaces." *Physical review letters* 119.10 (2017): 106802.
26. Ahmadi, Haddi, and Amin Khavasi. "Babinet-complementary structures for implementation of pseudospin-polarized waveguides." *Optics Express* 31.13 (2023): 21626-21640.
27. Zafari, Kazem, and Homayoon Oraizi. "Surface waveguide and y splitter enabled by complementary impedance surfaces." *Physical Review Applied* 13.6 (2020): 064025.
28. Ahmadi, H., Pasdari-Kia, M., Razmjooei, N., Ahmadi, Z., Zafari, K., Arik, K., & Oraizi, H. (2023). Quasi-Line Waves in Self-Inductive Impedance Surfaces.
29. Ahmadi, H., Zafari, K., Pasdari-Kia, M., Razmjooei, N., & Arik, K. (2023). Super ultra-wideband closed waveguide supporting one-way states.